\documentstyle[11pt,newpasp,twoside,epsf]{article}
\def\edcomment#1{\iffalse\marginpar{\raggedright\sl#1\/}\else\relax\fi}
\reversemarginpar

\begin{document}
\title{Dynamical Evolution of Solids Subject to the Drag
Force and the Self-Gravity of an Inhomogeneous and Marginally
Gravitationally Unstable Disk}
\author{Nader Haghighipour}
\affil{Dept. of Terrestrial Magnetism, Carnegie Institution
of Washington, 5241 Broad Branch Road NW, Washington, DC 20015
-1305, U.S.A.}

\begin{abstract}
The results of an extensive numerical study of the orbital dynamics
of small bodies ranging from micron-sized dust grains to 1 km objects
subject to gas drag and also the gravitational attraction of a 
non-uniform gaseous nebula are presented. The results indicate that
it is possible for small bodies to migrate rapidly toward the 
locations of the maxima of the gas density where the probabilities
of collisions and coagulations are enhanced. 
\end{abstract}

\section{Introduction}

It has recently been pointed out that a solar nebula massive enough
to form gas-giant planets through the core-accretion model is likely
gravitationally unstable (Pollack et al. 1996; Boss 2000; Inaba $\&$
Wetherill 2002). Such an unstable nebula is not entirely undesirable.
The alternative model of the giant planets formation, namely, 
the disk instability mechanism, suggests rapid formation of gas-giant 
planets followed by sedimentation of small solids at the locations of
spiral arms and clumps of a gravitationally unstable disk. It is,
therefore, fundamentally important to study the dynamical evolution
of solids in such an environment and in particular, the implications
for the collisional coagulation and growth processes.

A turbulence-free rotating gaseous nebula is at hydrostatic equilibrium
when the gravitational attraction of its central star is balanced
by a radial gradient in its pressure known as the pressure gradient 
(Figure 1). When the pressure gradient is positive, the velocity
of a gas molecule is greater than its local Keplerian velocity
(Eq. (1)). A solid in the gas, in this case, feels, effectively,
a larger acceleration along its orbit and, consequently, the increase
in its orbital angular momentum forces the solid to a larger orbit.
The opposite is true when the pressure gradient is negative.
\vskip -6pt
\begin{equation}
r{\omega_{\rm g}^2}\,=\,r{\omega_{\rm K}^2}\,+\,
{1\over {\rho_{\rm g}}}\,{{d{P_{\rm g}}}\over {dr}}
\qquad,\qquad
{\omega_{\rm K}^2}\,=\,{{GM}\over {r^3}}
\end{equation}
\vskip -1pt
\noindent
In this equation, $M$ is the mass of the central star, $G$ is the 
gravitational constant and ${P_{\rm g}},{\rho_{\rm g}}$ and
$\omega_{\rm g}$ represent the pressure, the density and the
angular velocity of the gas, respectively.

In a rotating gravitationally unstable disk, gas density
enhancements appear in the forms of spiral arms and clumps.
It is possible for the pressure of the gas

\begin{figure}
\vskip -160pt
\plotone{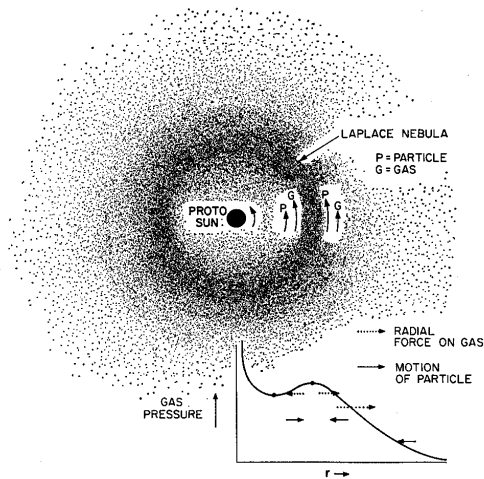}
\caption{Pressure gradient and the radial migration of solids
(Whipple 1972). The velocity of the gas slightly differs
from Keplerian circular (Eq.(1)).}
\end{figure}
\noindent
to have a radial
gradient in the vicinity of such density enhancements. In this
case, solids migrate toward the location of the maximum gas
density where the probabilities of their collisions and
coagulations are enhanced. In this paper, we study the dynamics
of solids that undergo such migrations while subject to gas drag
and the gravitational attraction of the nebula.

\section{The Model}

To focus attention on the rates of the in/outward migrations
of bodies and also their dependence on the solids and the
nebula's physical properties, an isothermal nebula of pure
molecular hydrogen with a sun-like star at its center and
with an azimuthally symmetric density given by Figure 2 is
considered here. The small solid bodies are considered to be
non-interacting and their motions are restricted to the midplane
of the nebula. The gravitational force of the nebula is also
taken into account (Haghighipour $\&$
Boss 2002, astro-ph/0207345).
In such a nebula, $\lambda$(mean free path
in cm)= $4.72 \times {10^{-9}} {\rho_{\rm g}^{-1}}$ g cm$^{-3}$,
$\sigma$(collisional cross section)= $2 \times {10^{-15}}$ cm$^2$
and the drag force of the gas is given by 
\begin{equation}
{{\bf F}_{\rm drag}}\,=\,-\,{4\over 3}\,\pi\,{R_{\rm d}^2}\,
{\rho_{\rm g}}\,\Bigl[(1-f){{\bar v}_{\rm th}}\,+\,
{3\over 8}f{C_{\rm D}}{v_{\rm rel}}\Bigr]{{\bf V}_{\rm rel}}.
\end{equation}
\noindent
In this equation, $f={R_{\rm d}}/({R_{\rm d}}+\lambda)$
(Supulver $\&$ Lin 2000) and $R_{\rm d}$ and ${\bar v}_{\rm th}$
represent the radius of the solid and the mean thermal velocity
of the gas molecules, respectively. The quantity $C_{\rm D}$
in equation (2) is the drag coefficient and is approximately
equal to $24/{R{\rm e}}$ for ${R{\rm e}}<1,\,24/{{R{\rm e}}^{0.6}}$
for $1<R{\rm e}<800$ and 0.44 for 
$R{\rm e}>800$ where
${R{\rm e}}=6({\rho_{\rm g}}{R_{\rm d}}\sigma/{m_0})
({v_{\rm rel}}/{{\bar v}_{\rm th}})$ is the gas Reynolds number and
$m_0$ represents the molecular mass of the gas.

\begin{figure}
\plottwo{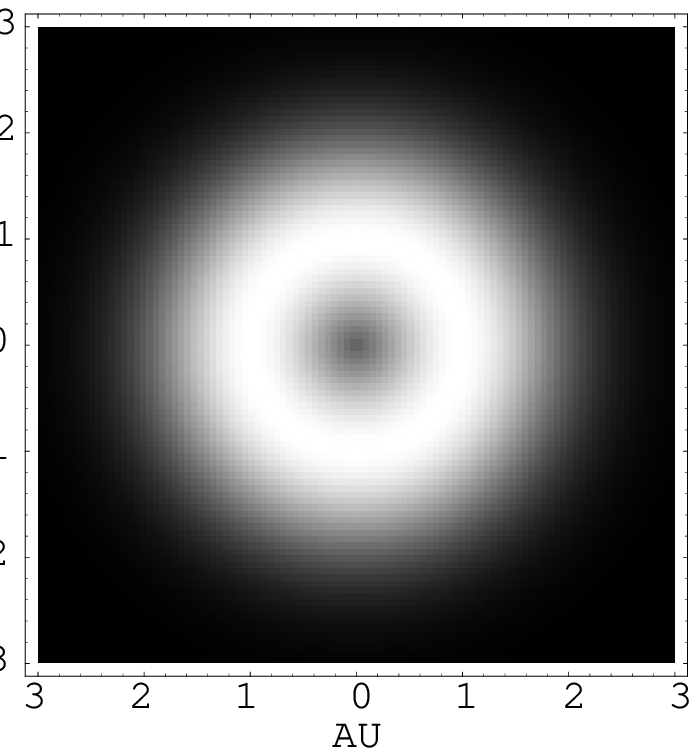}{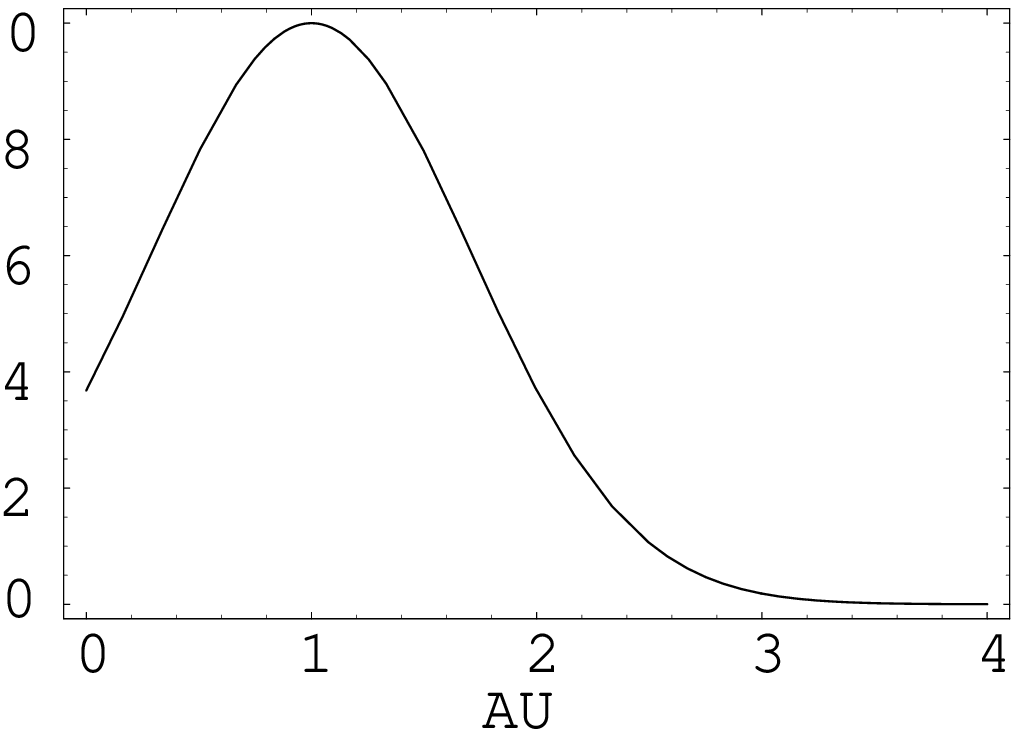}
\vskip -50pt
\caption{The density of the nebula in this model is
considered to have a general functional form of
${\rho_0}{\rm Exp}[-\alpha(r({\rm AU})-1)^2]$.
This figure shows ${\rho_{\rm g}}(r)$ for ${\rho_0}={10^{-9}}$
g cm$^{-3}$ and $\alpha=1\,{({\rm AU})^{-1}}$.}
\end{figure}

\section{Numerical Analysis}

The equations of motions of solids with sizes ranging from 1 micron
to 1 km have been numerically integrated for different values of the
solids densities and the gas temperature. Figure 3 shows the migration
of solids with densities equal to 2 and 5 g cm$^{-3}$ and radii of
10 and 100 cm. The temperature of the gas is constant at 1000 K. 
As shown here, solids with radii from 10 to 100 cm undergo rapid
in/outward migrations toward the location of maximum density. The time
of migration within a 1 AU neighborhood of $r=1$ (AU) is less then
1000 years, a time that is comparable with the time of giant planets
formation as suggested by the disk instability model (Boss 2000).

The rate of migration is also affected by changing the densities of
the solids. We studied the migration of solids for different
values of the solids densities. The results indicate that for cm-sized
objects, the rate of in/outward migrations increase with increasing 
the solids densities. For m-sized and larger objects, on the other hand,
increasing the solids densities results in an increase in the rate
of inward migration but a lower rate for migrating outward 
(Figure 3). For a detailed analytical analysis on this see
Haghighipour $\&$ Boss (2002, astro-ph/0207345).

We also studied the effect of changing the temperature
of the gas on the migration of solids. In general, increasing
the temperature resulted in decreasing the rate of migration 
(Figure 4). We integrated the motions of solids with a variety
of radii and densities for the gas temperature ranging from
25 K to 1000 K. Detailed analysis shows that an increase in
the temperature of the gas will result in an increase in the
mean thermal velocity of its molecules which in turn results
in a decrease in the magnitude of the relative velocity of a 
solid. That means, the effect of the drag force in pushing 
a solid body in/outward weakens with increasing the gas
temperature.

\section{Conclusions}

In general, the rate of the migration of a solid in a turbulence-free
gaseous nebula in the presence of gas drag and the gravitational force
of the nebula varies with the solid's mass and size and
also with the gas
density and temperature. The results of this study indicate
that it is, indeed, possible for solids within certain ranges of size
and density to migrate rapidly to the locations of the 
maximum values of the gas density. Given the likelihood that the
solar nebula was marginally-gravitationally unstable, the
processes studied here may have enhanced the growth rates
of solid planetesimals.

This work is partially supported by the NASA Origin of the Solar System
Program under grant NAG5-10547 and by the NASA Astrobiology
Institute under the grant NCC2-1056.

\begin{figure}
\plotone{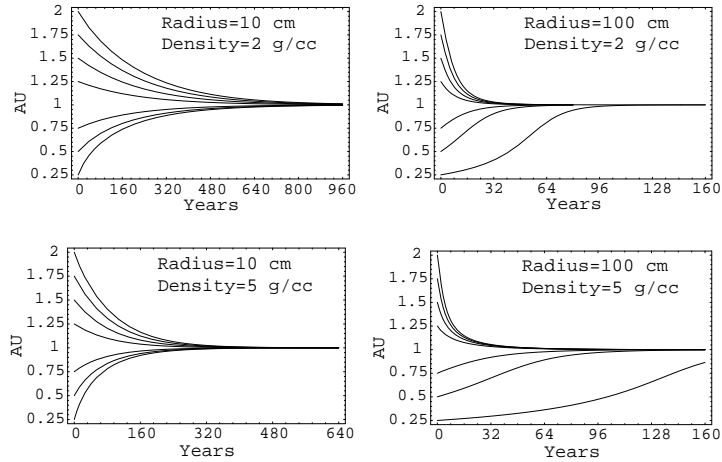}
\vskip -5in
\caption{Rapid migration of 10 cm and 100 cm objects with
densities equal to 2 and 5 g cm$^{-3}$ in an isothermal
solar nebula at 1000 K.}
\end{figure}

\vskip 4pt
\centerline{
\epsfbox{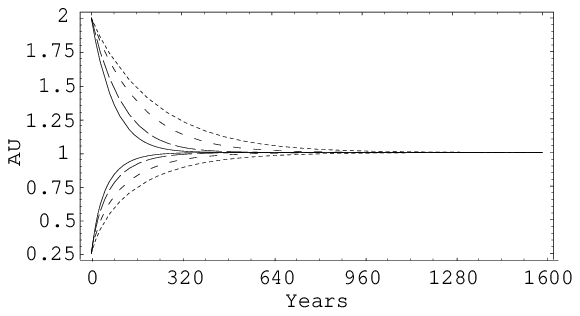}
\epsfbox{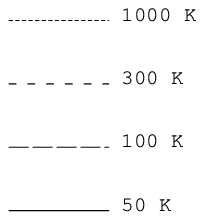}}
\vskip -80pt
{\rightskip 0.3in 
Figure$\,$4. \hskip 13pt Radial migration of a 10 cm-sized object with
a density of \par
2 g cm$^{-3}$ for four different values
of the gas temperature.\par}

\end{document}